**Leveraging mechanical resonances for the selection of promising materials in complex phase spaces**

Christopher A. Mizzi*, Osman El-Atwani, Tannor T.J. Munroe, Saryu Fensin, Boris Maiorov**

C. A. Mizzi
National High Magnetic Field Laboratory, Los Alamos National Laboratory, Los Alamos, New Mexico 87507, USA
Email: *mizzi@lanl.gov (corresponding)

O. El-Atwani
Pacific Northwest National Laboratory, Richland, Washington 99354, USA
Email: osman.elatwani@pnnl.gov

T.T.J. Munroe
National High Magnetic Field Laboratory, Los Alamos National Laboratory, Los Alamos, New Mexico 87507, USA
Email: tmunroe@lanl.gov

S. Fensin
Center for Integrated Nanotechnologies, Los Alamos National Laboratory, Los Alamos, New Mexico 87507, USA
Email: saryuj@lanl.gov

B. Maiorov
National High Magnetic Field Laboratory, Los Alamos National Laboratory, Los Alamos, New Mexico 87507, USA
Email: **maiorov@lanl.gov (corresponding)

**Funding**:
Los Alamos National Laboratory, Laboratory Directed Research and Development program project numbers 20220727ER and 20240225ER.

National Science Foundation Cooperative Agreement No. DMR-2128556, the State of Florida, and the Department of Energy.

U.S. Department of Energy, Advanced Research Projects Agency-Energy under contract DE-AR0001541

U.S. Department of Energy, Fusion Energy Sciences program under project No. 82396A, contract No. AT2030110.

**Abstract**

The "high-entropy" paradigm has been applied to a central challenge in materials science, the design of new functional materials with enhanced performance for targeted applications, with some notable successes over the last twenty years. However, the immensity of the high-entropy design space remains a major impediment to discovering optimal compositions with tailored microstructures. Suites of high-throughput computational tools have been developed to address this problem, but there is a compelling need to inform these models with fast, economical, non-destructive, and versatile experimental guidance. In this work, we demonstrate mechanical resonance measurements can fulfill this need. Mechanical resonance measurements enable the rapid, non-destructive assessment of materials created by novel syntheses and/or processes and provide high-accuracy determinations of elastic constants to directly benchmark models. We exemplify these capabilities on W-Ta-Cr-V-Hf and Mo-Nb-Ti-V-Zr refractory high-entropy alloys and suggest methodologies for the wider adoption and application of mechanical resonance measurements.

## I. Introduction

High-entropy materials (*1, 2*) have generated much interest owing to their potential for exceptional performance, tunability, and unique combinations of materials properties (*3, 4*). For example, some high-entropy materials combine strength and ductility which is especially attractive for mechanical applications (*5, 6*), while others exhibit promising irradiation resistance for utilization in nuclear reactors (*7, 8*). However, there are significant obstacles in optimizing the chemical composition and microstructure of high-entropy materials for tailored applications stemming from the immense size of the high-entropy design space (*9*). Two main challenges arise in all high-entropy material development and advancing their technology readiness level for industrial application: (1) Design considerations must balance mixing entropy and enthalpy along with microstructure. While maximizing entropy can improve properties and functionality, compositions need to optimize enthalpy to achieve homogenous materials without unwanted phases that can deteriorate performance; (2) Successful manufacturability and cost-effective fabrication of high-entropy materials that meets design requirements are composition-dependent. Every high-entropy material, and in some cases every composition, should have its own fabrication approach which, in general, requires large efforts in time-intensive parameter optimization and fabrication assessment. These two challenges motivate the development of high-throughput processes for the design, fabrication, and testing of high-entropy materials.

Specifically, there is a pressing need to complement large efforts in high-throughput, computational materials design of high-entropy systems currently underway (*e.g.*, Ref. (*10, 11*)), with rapid, non-destructive, and quantitative property assessments to down-select promising compositions, manufacturing methods, and/or thermomechanical processes for further study and provide experimental inputs to facilitate model development. Some work has started in this vein (*12-15*), especially for high-entropy thin films (*16*), but techniques which are compatible with bulk samples are still required.

A promising technique to address these needs is resonant ultrasound spectroscopy (RUS). RUS measures the frequencies and widths of the mechanical resonances of a sample (*17*) to characterize elastic constants (*18-20*) and internal friction (*21, 22*). Elastic constants are directly relevant for mechanical and structural applications, and provide important experimental insights into bonding, anharmonicity, and other aspects of the fundamental materials science of high-entropy materials which naturally connect with theory and modelling (*23*). Internal friction is useful to discern the type and number of defects present in a sample, and thereby quantify sample quality, because different types of defects have unique internal friction signatures.

While RUS has been used to study elasticity in select high-entropy materials, mostly derivatives of Co-Cr-Fe-Mn-Ni high-entropy alloys (*e.g.*, Ref. (*5, 24-27*)) with some work on refractory high-entropy alloys (*28-30*) and other systems (*31, 32*), we posit RUS has untapped potential in materials design workflows, especially as a means to rapidly, and non-destructively, determine sample quality in combination with other, more common techniques. Mechanical resonance measurements are fast (~minutes), accurate (absolute accuracy ~0.1%), sensitive (relative sensitivity ~ppm), and non-destructive (strain ~$10^{-6}$). They are particularly powerful to rapidly down-select high-entropy materials because there are no restrictions on the sample size or geometry (*i.e.*, mechanical resonance spectra can be obtained on as-manufactured materials) and the measurements can be performed with a simple, table-top apparatus (*e.g.*, see Ref. (*33*)). While no single technique can be used to explain performance improvements in new materials derived from novel syntheses and/or thermo-mechanical processes, the aforementioned combination of characteristics makes RUS ideal for high throughput and rapid, go/no-go classifications. After this first step of identifying promising candidates with RUS, one can apply intensive and specialized characterization and analytical techniques, often involving costly, time-consuming, and destructive sample preparation, that usually require chemistry and composition-dependent optimizations. One possible high-entropy design workflow incorporating RUS is provided in Fig. 1.

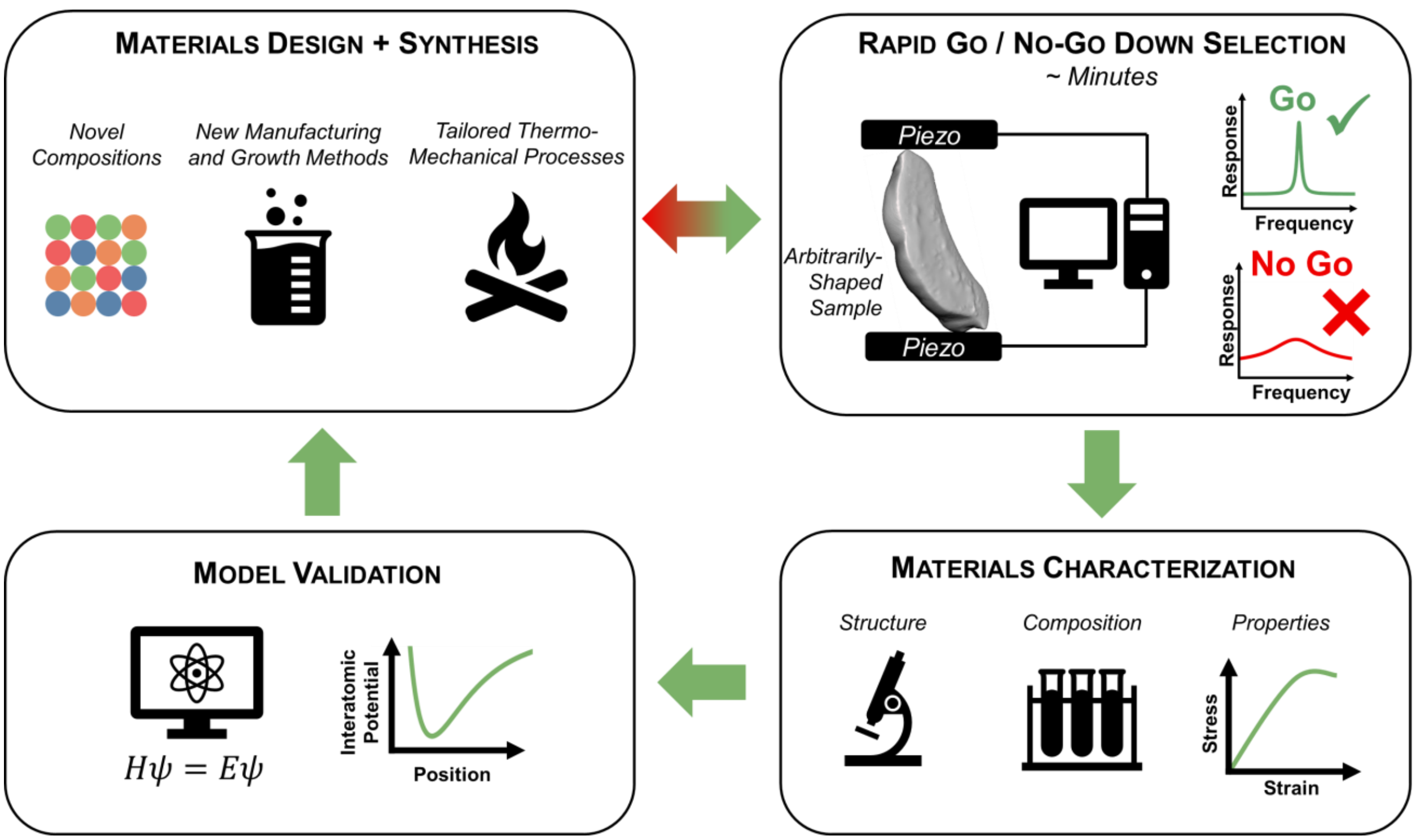

**Figure 1**. Materials design workflow incorporating resonant ultrasound spectroscopy for rapid down-selection, elastic property determination, and model validation. Owing to the speed with which mechanical resonance spectra can be acquired and compatibility with arbitrarily-shaped samples, RUS is promising for go, no-go rapid down selection of new materials grown with novel approaches and/or subjected to new thermo-mechanical processes. Once candidates have been identified via an iterative design/synthesis/down-selection process, other characterization techniques can be used to determine structure, composition, and properties which in turn can be compared with theoretical predictions to inform the next iteration of materials design/synthesis.

As summarized in Fig. 2, and elaborated upon below, the utility of RUS is rooted in the versatility of internal friction and elastic constants. To leverage these capabilities for down-selection, it is necessary to develop metrics derived from the widths and frequencies of mechanical resonances. We first focus on internal friction. Internal friction can be used to rapidly identify promising new compounds and/or processing conditions via the ultrasonic quality factor ($Q$), defined as the ratio of the mechanical resonant frequency ($f$) to the width of the resonance ($w$) (*i.e.*, $Q = \frac{f}{\mathrm{w}}$ (*19*)). Simply put, higher ultrasonic quality factors indicate higher quality samples. This is because the ultrasonic quality factor is inversely proportional to ultrasonic attenuation in the limit of low dissipation (*21, 22*) making it is very sensitive to the presence of point/line defects (Fig. 2, Panel I), cracks/pores (*34*) (Fig. 2, Panel II), and inhomogeneities (*e.g.*, internal stress and secondary phases). As such, it is a useful proxy for sample quality (*21, 22*). Ultrasonic quality factors can diagnose sources of low sample quality because different defects have unique ultrasonic quality factor signatures. For instance, high defect density tends to increase dissipation (Fig. 2, Panel I) and cracks introduce non-linear responses (Fig. 2, Panel II). The temperature dependence of the ultrasonic quality factor can also provide insights into the types of defects and their activation energies (*21, 22*), which can constrain *ab initio* defect calculations (*e.g.*, Ref. (*35*)).

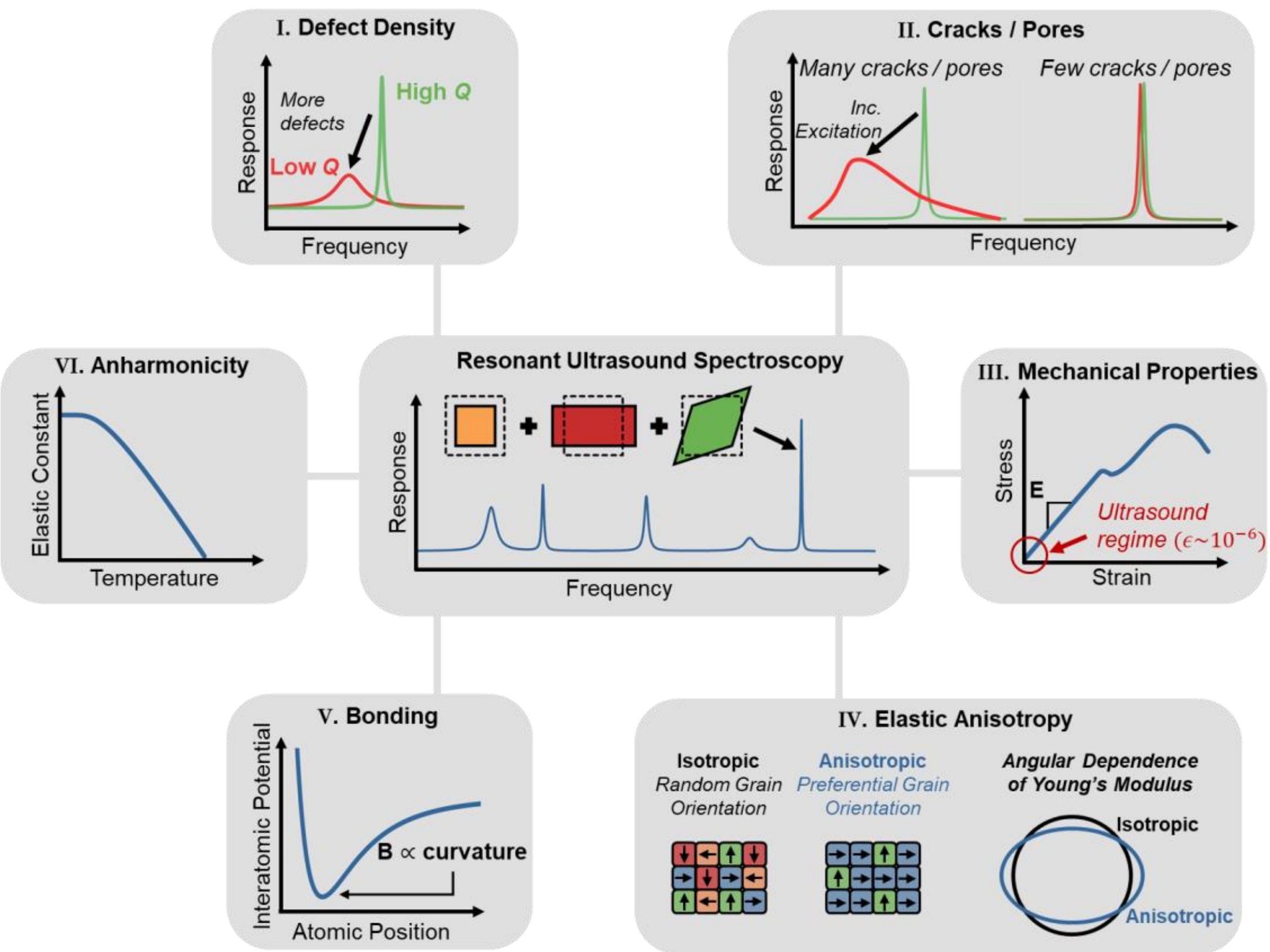

**Figure 2**. Examples of information that can be extracted from mechanical resonance spectra. *Center panel*: Each resonant frequency contains information on a weighted average of elastic constants. Resonance widths probe internal friction. *Panels I-VI:* Different aspects of sample quality and materials properties that can be determined from mechanical resonance spectra. The panels are arranged by the relative ease with which the information can be obtained (*e.g.*, *Panels I* and *II* require no sample preparation or analysis whereas *Panel VI* requires specialized equipment and analysis). Additional details on *Panels I-VI* are provided in the body of the paper.

The other primary measurable from a mechanical resonance experiment, the resonant frequency spectrum of the sample, depends on the sample geometry, density, and elastic constant tensor (*18, 19*). For samples with a known geometry and density, one can accurately, precisely, and efficiently determine the entire elastic constant tensor from a single resonant frequency spectrum of a single sample (*36*). Elastic constants determined by RUS correspond to the adiabatic elastic response (*33*) to ultralow (~$10^{-6}$) strains (*37*), as shown in Fig. 2, Panel III. While efficient inversion from resonant frequencies to elastic constants has conventionally relied upon samples with simple geometries (spheres, cylinders, and rectangular parallelepipeds (*38-41*)), recent improvements have made rapid and precise elastic constant determination on arbitrarily-shaped samples more tenable (*42, 43*). Since elastic constants describe the extent to which materials resist elastic deformation (Fig. 2, Panel III), they provide important information for mechanical performance and structural applications (*44-47*). For example, the shear modulus plays an integral role in many strengthening mechanisms (*3*) and the ratio of the bulk to shear moduli is related to intrinsic ductility (*44*) and hardness (*46, 47*). Compared to other mechanical property characterization approaches where only one of these properties is measured at a time, RUS determines the entire elastic constant tensor from a single measurement on a single sample. This makes RUS especially well-suited to detect elastic anisotropies (Fig. 2, Panel IV) affecting mechanical performance (*e.g.,* from processing-induced texture (*48*)) as well as properties which depend on multiple elastic constants (*e.g.*, ductility (*44*)). Beyond applications, and especially relevant for high-entropy materials design, elastic constants also contain rich information about the fundamental physics of materials. As second derivatives of the free energy (*49*), elastic constants provide microscopic insights about the nature of bonding between atoms (*50-53*) (Fig. 2, Panel V), and their temperature dependence (*54*) measures phonon softening and anharmonicity (*23, 49*) (Fig. 2, Panel VI). These connections to the free energy mean elastic constants naturally bridge experiment and theory, making them prime candidates to serve as a benchmark for materials design, model development, and validation.

In this paper, we describe how RUS can be incorporated in a materials design workflow to rapidly and reliably identify promising novel high-entropy compositions, manufacturing methods, and thermomechanical processes for further study, characterize elasticity, and benchmark theoretical models. These capabilities are demonstrated with measurements on two families of refractory high-entropy alloys (HEA): W-Ta-Cr-V-Hf and Mo-Nb-Ti-V-Zr. The former alloy was recently presented as a candidate fusion refractory material with exceptional radiation tolerance to defect evolution and enhanced morphology control (*55, 56*). The latter is being explored for high-temperature structural applications (*16, 57, 58*). First, we use W-Ta-Cr-V-Hf samples to show how mechanical resonances can be used to quickly compare manufacturing methods and identify sensitivities to processing conditions. Then, we use Mo-Nb-Ti-V-Zr samples to exemplify RUS-enabled down-selection, elastic constant determination, and theoretical model comparison. By leveraging the unique capability of RUS to simultaneously examine how strength and ductility evolve with composition in Mo-Nb-Ti-V-Zr samples, we observe strength exhibits a significantly larger compositional dependence than ductility and evolves non-monotonically with density. Lastly, through a comparison with a range of theoretical predictions, we find elastic constants estimated using the rule of mixtures provide a qualitative understanding of these compositional trends that surpasses existing *ab initio* predictions, although none of the investigated models provide quantitative agreement with experiment.

## II. Results

### A. *Rapid Down Selection and Manufacturing/Processing Assessment: W-Ta-Cr-V-Hf and Mo-Nb-Ti-V-Zr*

To demonstrate the utility of mechanical resonances as a rapid probe of sample quality, we show how ultrasonic quality factor measurements on as-manufactured, arbitrarily-shaped W-Ta-Cr-V-Hf samples can quickly identify differences, and challenges, arising from growth methods and processing conditions. For illustrative purposes we focus on 20W-35Ta-5Cr-35V-5Hf, with similar results for multiple compositions within the W-Ta-Cr-V-Hf family (see SI). Figure 3 compares the mechanical resonances and ultrasonic quality factors of un-machined hot-pressed and arc-melted 20W-35Ta-5Cr-35V-5Hf samples. The mechanical resonances of the hot-pressed samples are substantially broader than those of arc-melted samples with the same composition, yielding ultrasonic quality factors that are nearly a factor of 10 smaller. Typically, $Q \gtrsim 10^4$ are ideal, but $Q{\sim}10^3$ can still allow for reasonable elastic constant determination. This measurement, which took ~1 minute, indicates the arc-melted samples are significantly more homogeneous and possess a lower defect concentration than their hot-pressed counterparts. Neither material showed non-linear responses indicating minimal cracks/pores, but density measurements of the arc-melted (13.66(2) g/cm$^3$) and hot-pressed samples (10.84(2) g/cm$^3$) suggest low quality factors in the hot-pressed sample is likely due to densification challenges in pressing (expected density is 13.3(2) g/cm$^3$ from a rule of mixtures estimate).

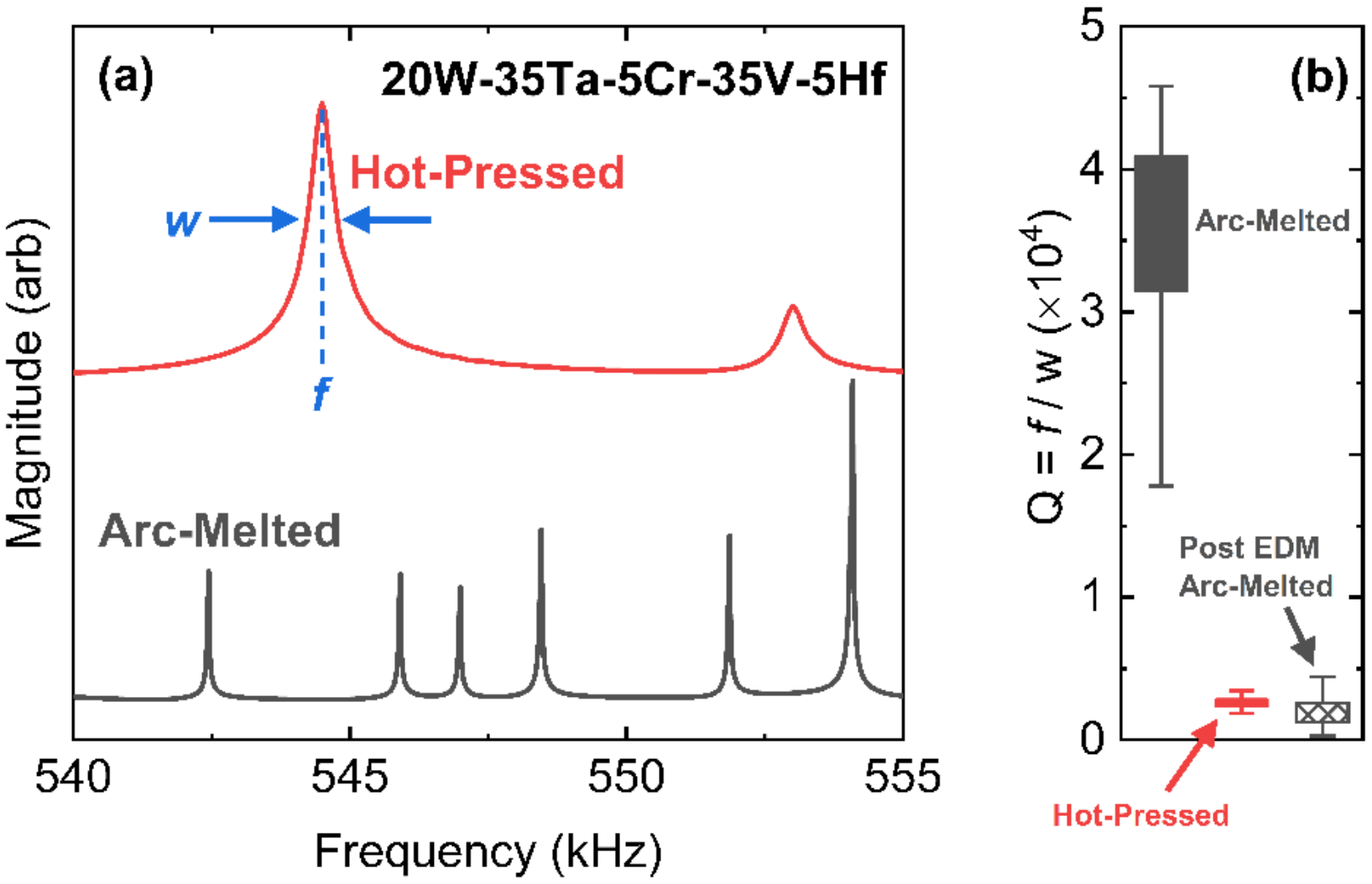

**Figure 3**. Rapid, non-destructive assessment from ultrasonic resonances compatible with samples of arbitrary shape and size. (a) Screening of 20W-35Ta-5Cr-35V-5Hf refractory high-entropy alloy samples synthesized with arc melting and hot-pressing. Spectra have been vertically offset for clarity. (b) The hot-pressed sample possesses much smaller ultrasonic quality factors ($Q$) indicative of increased ultrasonic attenuation from inhomogeneities and a higher concentration of defects.

Electrical discharge machining (EDM) resulted in substantial surface damage in an arc-melted 20W-35Ta-5Cr-35V-5Hf sample. Median ultrasonic quality factors are $3.8\times10^4$, $2.6\times10^3$, and $1.9\times10^3$ for the arc-melted, hot-pressed, and arc-melted post EDM samples, respectively. The box plots show ultrasonic quality factors of the first 60 resonances with color regions and whiskers denoting the 25th-75th percentiles and maxima/minima, respectively.

Next, we prepared the arc-melted W-Ta-Cr-V-Hf samples with EDM into shapes for conventional RUS (*18, 19*) to quantify their elastic constant tensor. However, measuring the ultrasonic quality factors after EDM revealed a major processing sensitivity: EDM damaged the W-Ta-Cr-V-Hf samples causing the mechanical resonances to substantially broaden such that the ultrasonic quality factors of the machined arc-melted W-Ta-Cr-V-Hf samples became comparable to the un-machined, hot-pressed samples (Fig. 3). This measurement exemplifies how a damage assessment from specific processing conditions can be rapidly ascertained from comparing changes in the values of ultrasonic quality factors. To further understand the origin of the quality factor reduction with EDM, samples from the same as-grown, arc-melted piece were prepared with different specific surface areas (a measure of damage corresponding to the ratio of surface area exposed to EDM cutting and total volume (*59*)). The quality factors shift to lower values as the specific surface area is increased, indicating surface damage from the cutting process (such as a decrease in density in the near surface region caused by disorder) is likely responsible for the reduction in quality factors (Fig. 4). This interpretation is further supported by density measurements in the three samples which decreased from 13.66(2) g/cm$^3$ for the as-grown material to 13.20(2) g/cm$^3$ and 12.11(2) g/cm$^3$ in the larger and smaller samples cut with EDM, respectively. This result indicates that although arc-melted W-Ta-Cr-V-Hf samples are high quality as-grown, they are sensitive to damage from EDM sample preparation, and alternative approaches should be used to process such samples (*e.g.*, high temperature annealing post EDM).

While quantitative analyzes of quality factors can be complicated because quality factors are frequency-dependent (with higher frequencies probing smaller length scales) and symmetry-dependent (each resonance contains information on a linear combination of symmetrized deformations) (*19, 60*), these results demonstrate how quality factors of mechanical resonances provide actionable qualitative information for materials design. Mechanical resonances can be used to rapidly (~minutes) decide if samples warrant further, and more resource-intensive, characterization, additional/alternative processing (*e.g.*, anneals to relieve internal stress), or should simply be discarded. The fast turnaround time, low cost, and small footprint of this ultrasonic probe, means such measurements can also be performed at multiple points throughout the fabrication and processing history of a sample allowing high throughput down-selection and rapid sample preparation optimization.

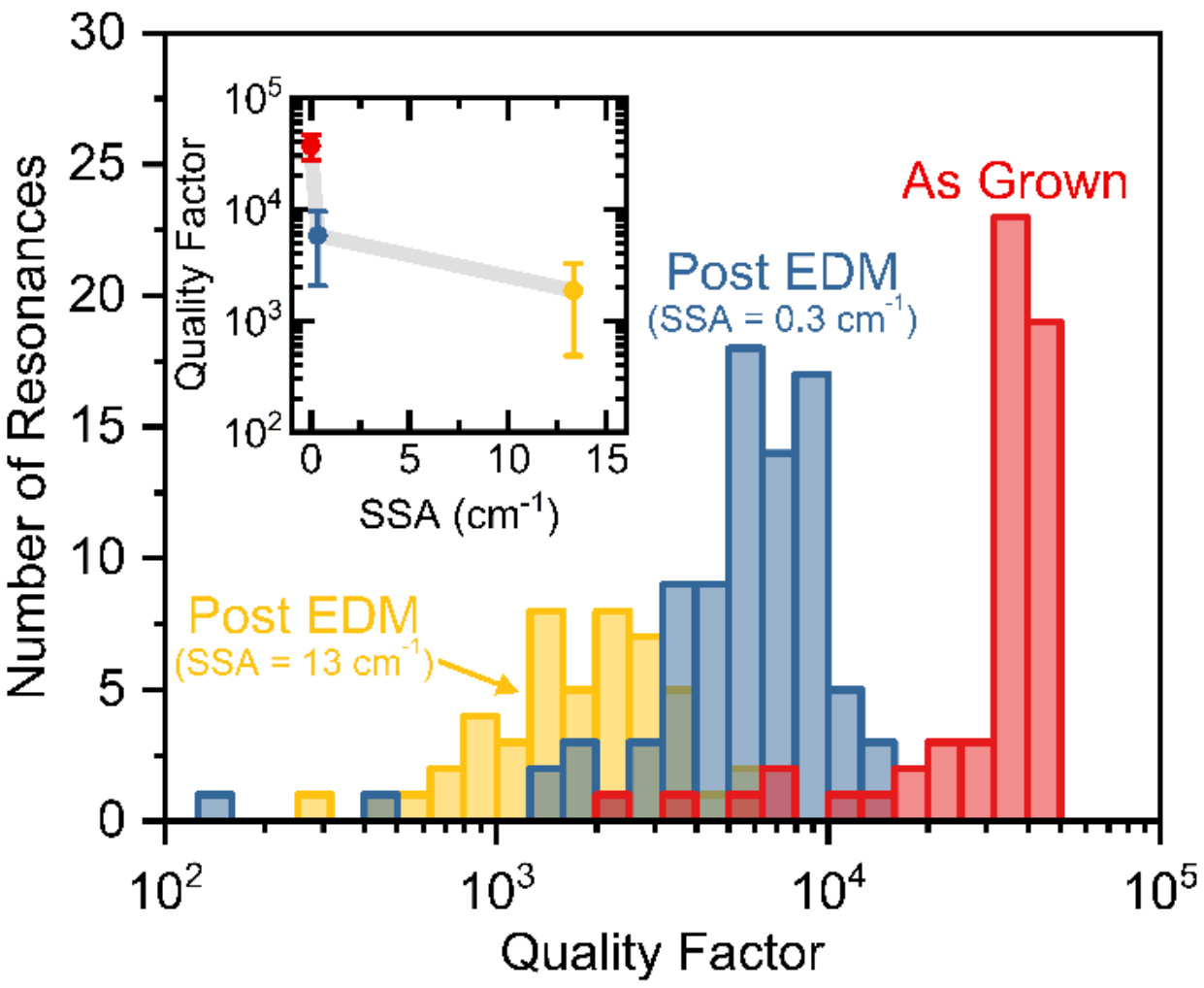

**Figure 4**. Effects of electrical discharge machining (EDM) on an arc-melted 20W-35Ta-5Cr-35V-5Hf sample. Histogram of quality factors for the as grown sample and two samples of different sizes after EDM cutting. The as grown sample exhibited high ultrasonic quality factors (median value of $3.8\times10^4$) which were reduced to median values of $5.6\times10^3$ and $1.9\times10^3$ for the larger and smaller samples, respectively, after EDM. (inset) Median quality factor and interquartile range for the three samples as a function of the specific surface area (SSA, ratio of cut area to volume) shows the reduction in the quality factor scales with SSA.

The negative effects of EDM on the arc-melted W-Ta-Cr-V-Hf precluded a more detailed elastic property investigation and comparisons with theoretical predictions using conventional RUS, but this was not the case for the samples from the Mo-Nb-Ti-V-Zr family of refractory HEA compounds. After EDM was used to cut Mo-Nb-Ti-V-Zr samples into cylinders, the samples possessed high ultrasonic quality factors ($\sim 10^4$) indicative of minimal EDM damage (Fig. 5). The distributions of the $Q$ values for different samples of a particular composition were also similar, indicating homogeneity.

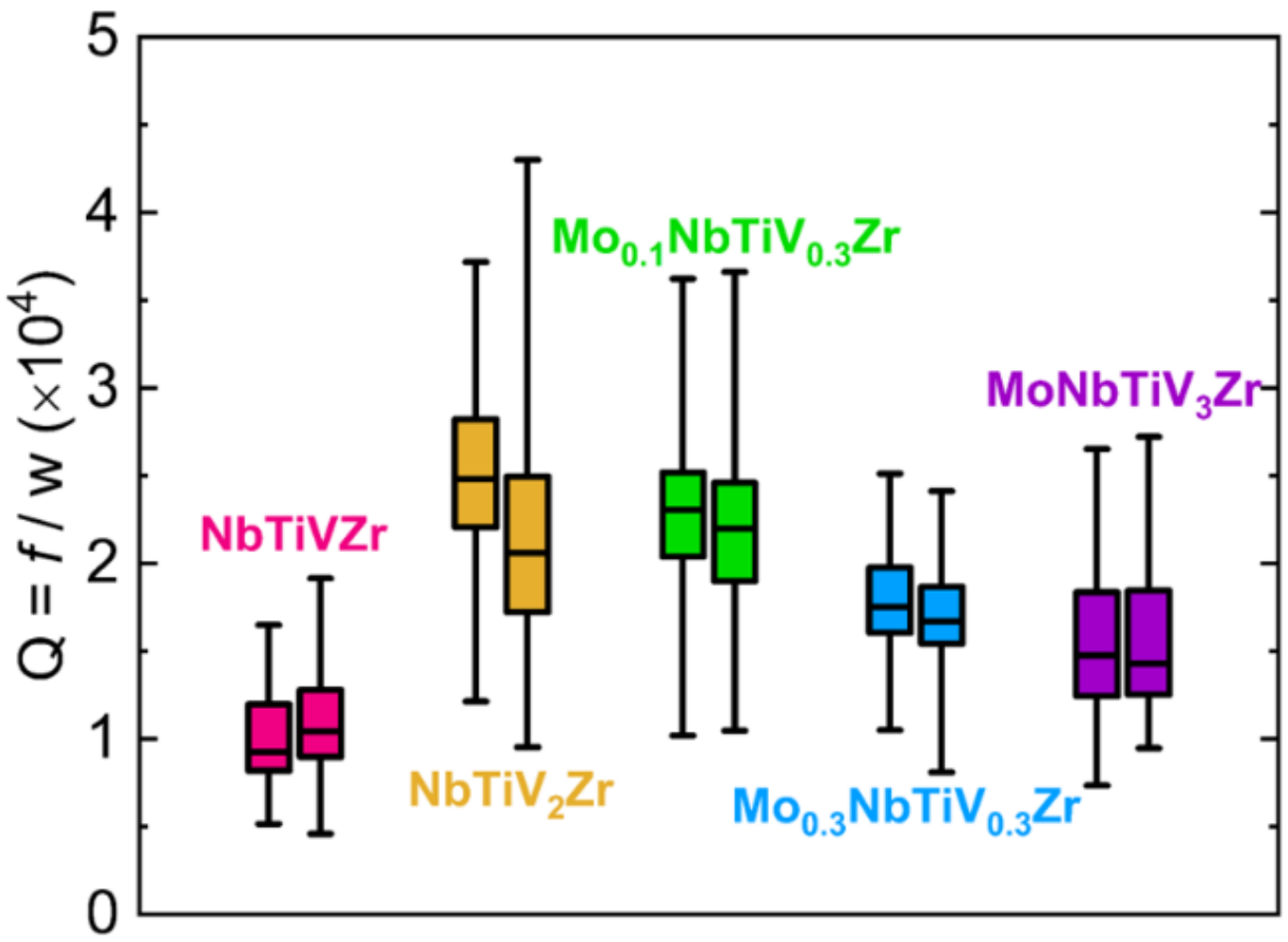

**Figure 5.** Boxplots of ultrasonic quality factor ($Q$) measured for two samples of each composition of Mo-Nb-Ti-V-Zr refractory high-entropy alloys studied in this work. All samples show high median ultrasonic quality factor values (solid line) and consistency between multiple samples of the same compound. Color regions denote the 25th-75th percentiles and whiskers give the maximum and minimum values from the first 60 resonances.

*B. Non-Destructive Elastic Property Determination and Model Validation: Mo-Nb-Ti-V-Zr*

Having down-selected the Mo-Nb-Ti-V-Zr samples for further study owing to their homogeneity and high quality (Fig. 5), we now examine their elastic properties in greater detail to understand the effects of composition on mechanical performance and exemplify using elastic constants determined from mechanical resonances for model validation. The Mo-Nb-Ti-V-Zr compositions investigated in this study were NbTiVZr, $NbTiV_2Zr$, $Mo_{0.1}NbTiV_{0.3}Zr$, $Mo_{0.3}NbTiV_{0.3}Zr$, and $MoNbTiV_3Zr$ to examine the relationship between density, strength, and ductility. Table I includes geometric densities and isotropic elastic constants determined with RUS at room temperature. To confirm homogeneity and reproducibility, two samples of $NbTiV_2Zr$, $Mo_{0.1}NbTiV_{0.3}Zr$, and $Mo_{0.3}NbTiV_{0.3}Zr$ were each measured and found to have densities and elastic constants in quantitative agreement (see SI). The values in Table I are averages of two samples for these three compositions. For all the samples studied here, isotropic elasticity was confirmed by reducing the symmetry of the elastic model in the RUS inversion (*48*) to quantify the extent of elastic anisotropy (see SI). Estimates of elastic Debye temperatures from the polycrystalline elastic constants (*61, 62*), corresponding to the long-wavelength phonon contributions to the total Debye temperature (see SI), are provided in Table I. Values from the literature are also included in Table I.

As composition is varied within the Mo-Nb-Ti-V-Zr family of compounds, our ultrasound measurements indicate the bulk and shear moduli change by ≈20%, while the Poisson's ratio is relatively constant, changing by only ≈2%. The similar Poisson's ratio of $\nu \approx 0.37$ and Pugh's ratio $\frac{B}{G} \approx 3.5$ for all compounds indicate the bonding (*51-53*) and intrinsic ductility (*44*) are similar and have a weaker compositional dependence. Comparatively, the shear modulus has a larger compositional dependence suggesting the strength will more prominently vary with composition (*3*). Experimental values for the Young's moduli of NbTiVZr and $NbTiV_2Zr$ obtained from compression testing (*57*) exhibit higher uncertainty than our ultrasound measurements, but fall within our uncertainties. This result confirms the elastic moduli determined here agree with those determined with mechanical testing. The Young's modulus extracted from nanoindentation measurements (*63*) on NbTiVZr is ≈14% higher than our ultrasound values and the literature compression testing values (*57*), possibly because of local differences in composition or homogenization.

| Material | Method / Reference | $\rho$ (g/cm$^3$) | B (GPa) | G (GPa) | E (GPa) | $\nu$ | B/G | $\theta_D^{elastic}$ (K) |
|---|---|---|---|---|---|---|---|---|
| NbTiVZr | Room T Exp. This work | 6.41(3) | 112(2) | 32.6(1) | 89.2(3) | 0.367(1) | 3.44(6) | 361.6(5) |
| | $T$ = 2K Exp. This work | | 114(2) | 33.9(1) | 92.6(2) | 0.365(2) | 3.37(7) | 368.6(5) |
| | RoM This work | 6.45(16) | 125(4) | 39.9(2) | 108.4(6) | 0.355(4) | 3.1(1) | 399(1) |
| | Compression Ref. (*57*) | 6.52 | | | 80(8) | | | |
| | Nanoindentation Ref. (*63*) | 6.35 | | | 102(3) | | | |
| | DFT Ref. (*63*) | 6.60 | 117.6 | 41.9 | 112.4 | 0.34 | 2.8 | |
| | DFT + Calphad Ref. (*64*) | | 130 | 28 | 78 | 0.40 | 4.6 | |
| | VCA Ref. (*65*) | | 133 | 31 | 86 | 0.39 | 4.3 | |
| | CPA Ref. (*66*) | | 117 | 44 | 118 | 0.33 | 2.7 | |
| | SQS Ref. (*66*) | | 130 | 20 | 57 | 0.43 | 6.5 | |
| | Ab initio Ref. (*67*) | | 118.6 | 45.70 | 121.1 | 0.330 | 2.6 | |
| $NbTiV_2Zr$ | Room T Exp. This work | 6.32(3) | 122(2) | 33.4(1) | 91.8(3) | 0.374(1) | 3.65(6) | 374.4(5) |
| | RoM This work | 6.40(14) | 129(4) | 41.0(3) | 111.3(7) | 0.357(4) | 3.2(1) | 413(1) |
| | Compression Ref. (*57*) | 6.34 | | | 98(10) | | | |
| | DFT + Calphad Ref. (*64*) | | 140 | 36 | 100 | 0.38 | 3.9 | |
| | VCA Ref. (*65*) | | 143 | 33 | 91 | 0.39 | 4.3 | |
| $Mo_{0.1}NbTiV_{0.3}Zr$ | Room T Exp. This work | 6.57(3) | 114(2) | 32.0(1) | 87.8(3) | 0.372(1) | 3.56(6) | 349.4(5) |
| | RoM This work | 6.59(20) | 123(5) | 40.4(9) | 109(2) | 0.353(6) | 3.1(1) | 391(1) |
| $Mo_{0.3}NbTiV_{0.3}Zr$ | Room T Exp. This work | 6.74(3) | 123(3) | 35.1(1) | 96.2(3) | 0.370(1) | 3.50(9) | 362.3(5) |

| | RoM<br>This work | 6.76(22) | 128(6) | 43(2) | 116(5) | 0.35(1) | 3.0(2) | 400(1) |
|---|---|---|---|---|---|---|---|---|
| $MoNbTiV_3Zr$ | Room T Exp.<br>This work | 6.82(3) | 137(4) | 39.5(1) | 108.1(4) | 0.368(1) | 3.5(1) | 397.9(5) |
| | RoM<br>This work | 6.88(20) | 145(7) | 49(4) | 133(9) | 0.35(1) | 2.9(3) | 441(2) |

**Table I.** Geometric density ($\rho$), isotropic polycrystalline bulk ($B$), shear ($G$), and Young's ($E$), moduli, Poisson's ratio ($\nu$), Pugh's ratio ($B/G$), and elastic Debye temperature ($\theta_D^{elastic}$) of the refractory high-entropy alloys studied in this work determined at room temperature, and 2K for NbTiVZr. Literature values are included for comparison. See SI for a detailed description of the rule of mixtures (RoM) calculations and references for experimental elastic constants and lattice parameters of each constituent element.

Next, we compare predicted elastic moduli across the Mo-Nb-Ti-V-Zr family using a variety of theoretical approaches (see Table I) with experimental results by plotting the bulk modulus as a function of the shear modulus for each compound (Fig. 6). Straight lines in Fig. 6 correspond to constant values of B/G and are related to the nature of bonding in the different compounds (*51-53*). Generally, larger values for Pugh's ratio indicate greater intrinsic ductility and more metallic bonding character, whereas smaller values are associated with brittleness and more covalent bonding character (*44, 51-53, 68, 69*). The crossover between these two regimes typically occurs around values of B/G≈1.7 in isotropic elastic systems, but will, in general, depend on the crystal- and micro-structure (*e.g.*, Ref. (*70*)).

The comparison in Fig. 6 indicates the theoretical predictions of the elastic constants for the Mo-Nb-Ti-V-Zr family of compounds exhibit large, non-systematic differences from experiment. Focusing on the case of NbTiVZr as an example (*63-67*): some predictions yield bulk moduli close to the measured value (*63, 66, 67*), while others yield shear moduli close to the measured value (*64, 65*), but none of the predictions are close to both the experimentally measured bulk and shear moduli. Moreover, the deviations in Poisson's and Pugh's ratios indicate that the methods used to model these compositions do not appropriately capture bonding (*51-53*) in the Mo-Nb-Ti-V-Zr family of compounds.

Because the differences between the magnitudes of the theoretically predicted and experimentally determined elastic properties do not appear systematic, it would be challenging to discern compositional trends in elastic properties from such calculations or use them to predict performance. Even for a fixed composition, the differences with experimental elastic properties can be substantial, leading to poor predictions of mechanical behavior. For example, predictions for Pugh's ratio (a proxy for intrinsic ductility (*44*) and hardness (*46, 47*)) in NbTiVZr span $2.5<B/G<6.5$, depending on the theoretical method.

We directly rule out the possibility that the discrepancy between the experimental and theoretical values of the elastic constants is temperature by measuring the elastic constants of NbTiVZr at 2K in addition to room temperature (Table I). *Ab initio* calculations are typically 0 K predictions and although most materials only stiffen on the order of a few percent on cooling from room temperature to 0 K (*49, 54*), some materials are known to exhibit large changes in elasticity with temperature (*e.g.*, $\delta$-Pu (*71*)), even without a phase transition (*49*). We find the shear and bulk moduli of NbTiVZr stiffened by ≈4% and ≈2%, respectively, from 300K to 2K which does not account for the discrepancy between the experimental and theoretical values shown in Fig. 6. It is worth noting the ratio of the bulk and shear moduli decreased with decreasing temperature, suggesting a reduction in intrinsic ductility of ≈2% as the temperature is decreased (*44*). This is consistent with the conventional trade-off between strength and ductility (*e.g.*, Ref. (*72*)).

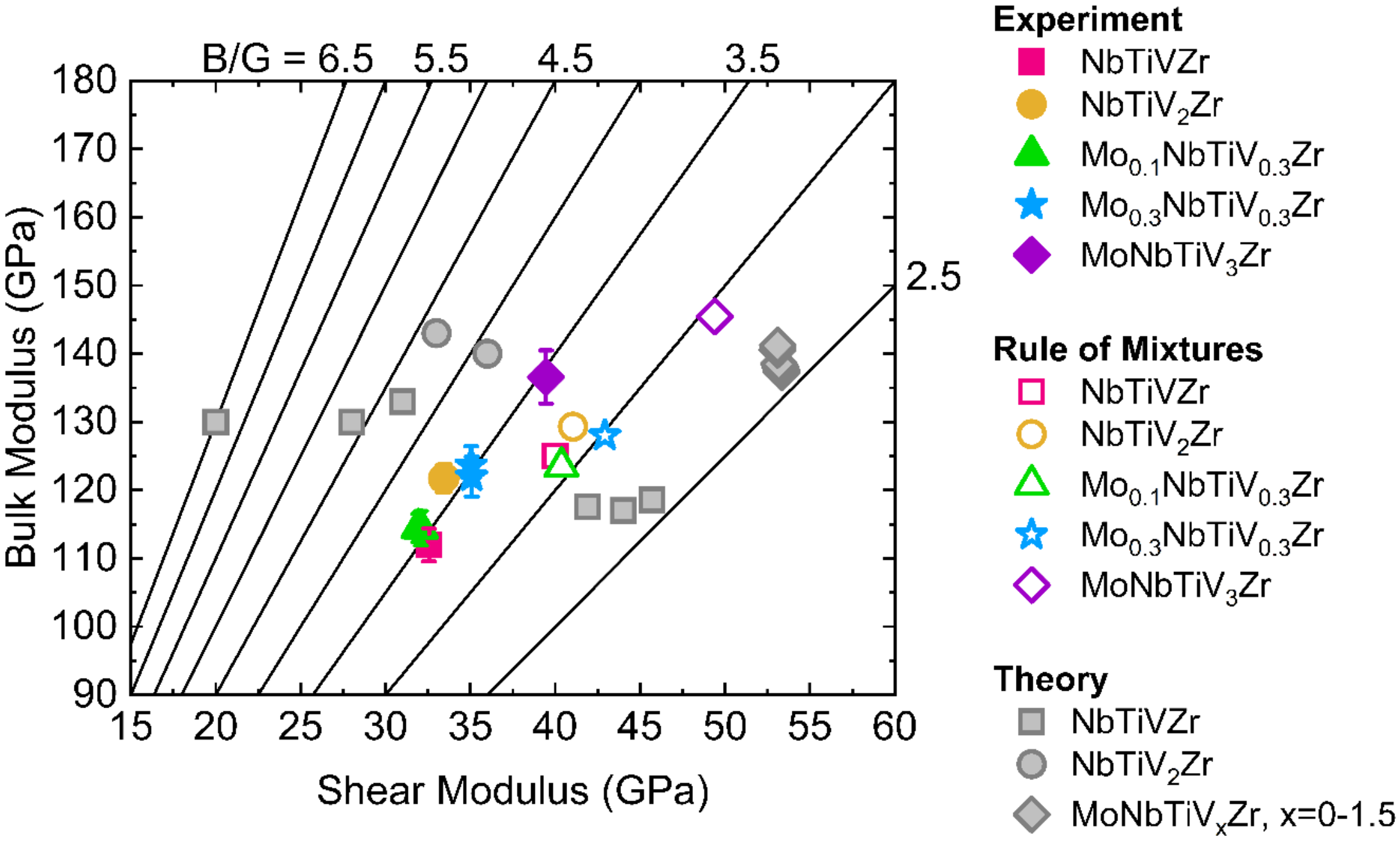

**Figure 6**. Bulk modulus as a function of shear modulus for compounds in the Mo-Nb-Ti-V-Zr family. Filled symbols correspond to ultrasound measurements reported in this work. Open symbols are calculated according to the rule of mixtures (Eq. 1-3, main text) using experimental values for the elastic constants (*73-76*) and lattice parameters (*77-80*) of the constituent elements. Grey symbols are theoretical calculations (*57, 63-67*). Solid lines correspond to different values of the ratio of the bulk modulus to the shear modulus.

Motivated by the poor agreement between theoretical predictions of elasticity in the Mo-Nb-Ti-V-Zr family and the measured values, we explored if simpler models could capture the compositional trends in the experimental elastic constants. We found a rule of mixtures prediction using experimental values for the elastic constants (*73-76*) and lattice parameters (*77-80*) of the constituent elements provides a better qualitative understanding of the evolution of elastic constants within this compositional series. As shown in Fig. 6, the rule of mixtures predictions captures compositional trends in Mo-Nb-Ti-V-Zr elastic properties, with quantitative differences. As such, we now focus on thorough comparisons between experimental elastic properties and rule of mixtures predictions to demonstrate the utility, and shortcomings, of the latter. In all cases, the rule of mixtures estimates for a physical property ($M_{RoM}$) were determined via

$$M_{RoM} = \frac{1}{2}\left(M_{RoM}^{lower} + M_{RoM}^{upper}\right). \quad (1)$$

$M_{RoM}^{upper}$ and $M_{RoM}^{lower}$ correspond to upper and lower bound rule of mixtures estimates calculated from

$$M_{RoM}^{upper} = \frac{\sum_i c_i \, V_i \, M_i}{\sum_i c_i \, V_i} \quad (2)$$

and

$$\frac{1}{M_{RoM}^{lower}} = \frac{\sum_i c_i \, V_i \frac{1}{M_i}}{\sum_i c_i \, V_i} \quad (3)$$

where experimental values of molar volume $V_i$, mole fraction $c_i$, and physical property $M_i$ (density, bulk modulus, or shear modulus) of element $i$ were used (*81*). In analogy to averaging methods to obtain isotropic elastic constants from single crystal elastic constants (*61, 62*), Eq. 1-3 represent Hill arithmetic (*82*), Voigt (*83*), and Reuss (*84*) averages, respectively. Additional information on the rule of mixtures calculations is provided in the SI.

Figure 7 demonstrates the measured densities track with rule of mixtures expectations. This indicates that the rule of mixtures is a good predictor for density in this class of refractory high-entropy alloys. The qualitative trend in the elastic moduli across the composition series is also captured well by the rule of mixtures predictions, but there are quantitative differences: on average, the rule of mixtures bulk moduli are ≈7% higher than measured values and the rule of mixtures shear moduli are ≈24% higher than measured values. Similarly, while Pugh's ratio estimated from the rule of mixtures quantitatively deviates from experiment (see Fig. 6), the rules of mixtures estimates qualitatively capture all the experimental compositional trends.

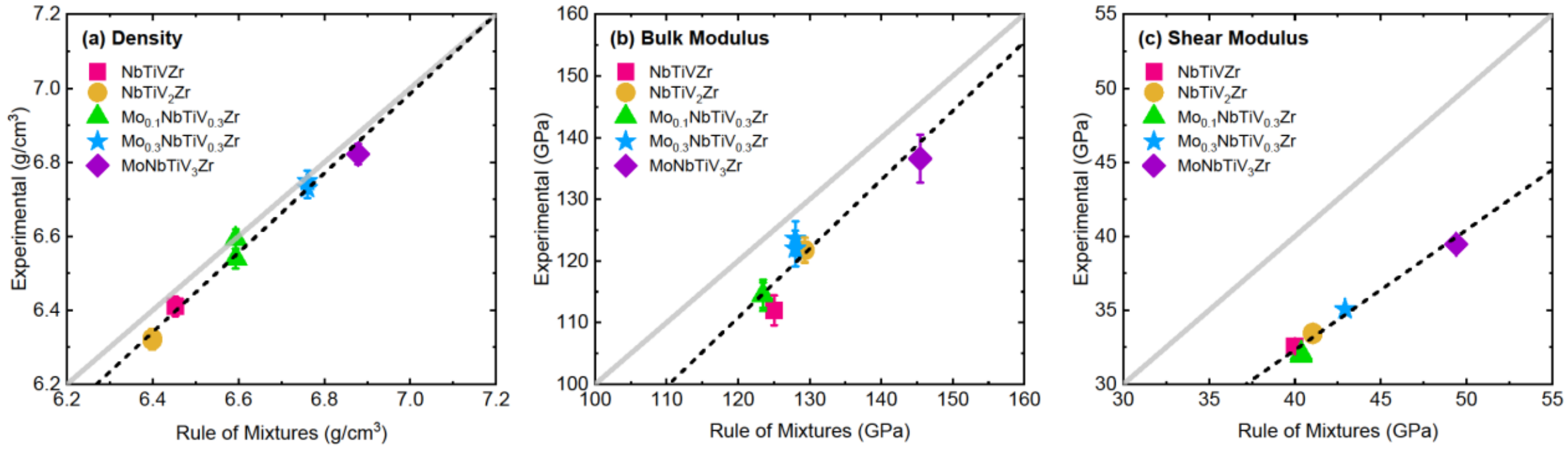

**Figure 7**. Measured (a) density, (b) bulk modulus, and (c) shear modulus plotted as a function of rule of mixtures predictions (Eq. 1-3, main text) for different stoichiometries (colors/symbols) in the Mo-Nb-Ti-V-Zr refractory high-entropy alloy family. Dashed black lines are linear fits and solid grey lines indicate perfect correspondence between experimental quantities and rule of mixtures predictions. The rule of mixtures captures the qualitative trends of all three properties, with quantitative discrepancies especially in the shear moduli.

## III. Discussion

The parameter space for designing, synthesizing, and processing new high-entropy materials is vast. High-throughput computational approaches are a critical component of efforts to efficiently identify promising novel high-entropy materials, but it is imperative to complement these approaches with experimental benchmarks and validation. Similarly, once new high-entropy materials have been fabricated and processed, there is a need to rapidly assess which samples/processes warrant further investigation. As demonstrated in this paper, resonant ultrasound-based approaches are an enticing addition to the high-entropy materials design process (Fig. 1) for both the rapid, non-destructive down-selection of novel high-entropy materials as well as model benchmarking and validation. The former is a consequence of the sensitivity of mechanical resonances to a variety of defects and the latter is derived from the range of applications-relevant information obtained from elastic constants (see Fig. 2). We posit there is great opportunity to pair the information encoded within mechanical resonances with other commonly used composition and structure probes to build machine learning models and create digital twins which can ultimately facilitate inverse materials design (*85, 86*).

The first system investigated here, bulk W-Ta-Cr-V-Hf refractory high-entropy alloys, showcased the ability of mechanical resonances to provide guidance on both synthesis and processing of high-entropy materials. Through examining changes in the ultrasonic quality factor as a function of fabrication method, we identified substantial differences between the hot-pressed and arc-melted samples likely due to a combination of densification challenges and differences in the concentration of defects. This assessment was performed in minutes, required minimal post-measurement analysis, and allowed us to down-select the arc-melted samples for further investigation. However, a comparably fast assessment of the arc-melted samples after cutting with EDM uncovered a high-degree of processing sensitivity, likely owing to surface damage, which precluded further analysis and identified the need for more research into the synthesis and processing of these alloys.

After selecting the Mo-Nb-Ti-V-Zr family of compounds as promising for additional investigation based on their ultrasonic quality factors, we determined the elastic constants in multiple Mo-Nb-Ti-V-Zr alloys. A key observation from measuring a range of compositions in the Mo-Nb-Ti-V-Zr family of refractory HEA compounds is the discrepancy between the experimental and predicted elastic moduli from a variety of theoretical calculations (*57, 63-67*). These differences are not systematic, making the source of error difficult to identify and account for *a priori*. Because of the role of the shear modulus in many strengthening mechanisms (*3*) and the effects of the bulk and shear moduli on intrinsic ductility (*44*) and hardness (*46, 47*), such differences have important implications for applications and exemplify the need for the integration of experimental input into materials design frameworks to benchmark theoretical predictions.

Furthermore, our measurements demonstrate an ability to improve strength across the Mo-Nb-Ti-V-Zr compounds without substantially impacting ductility and density: the shear modulus changes by ≈20% across the compositional series whereas Pugh's ratio and density only change by ≈6% and ≈8%, respectively. This concurrent assessment of multiple mechanical properties is enabled by the unique capability of RUS to simultaneously determine all elastic constants.

Prior studies which compared the rule of mixtures and experimental elastic constants in other high-entropy systems concluded that the rule of mixtures provided useful estimates of elastic constants (*29*). Compared with first principles calculations, we found the rule of mixtures estimates also quantitatively differed from experimental values but provided a good qualitative understanding of the systematic compositional dependence of elastic properties in the Mo-Nb-Ti-V-Zr family of materials. The latter observation is in agreement with Ref. (*29*) and suggests rule of mixtures estimates are useful first order approximations to elastic properties in high-entropy alloys. To further explore the rule of mixtures comparison, Fig. 8 shows how two elastic properties relevant for the mechanical behavior of materials, the shear modulus (strength (*3*)) and the ratio of the bulk and shear moduli (intrinsic ductility (*44*) and hardness (*46, 47*)) change with density within the Mo-Nb-Ti-V-Zr system. The ratio of the bulk and shear moduli decreases as a function of density, suggesting a decrease in ductility. In contrast, the shear moduli exhibit non-monotonic changes with density (similar behavior for bulk moduli, see SI). As shown in Fig. 8, both trends are captured by the rule of mixtures.

The non-monotonic change in the shear modulus as a function of density bears additional comment as this result appears to buck convention: often, one expects elastic constants to scale with density (*87*). The non-monotonic relationship between shear modulus and density is a direct consequence of elastic properties of HEAs being a weighted average over the properties of the individual elemental constituents. Since our experiments establish it is qualitatively correct to interpret elastic properties in this class of HEAs via the rule of mixtures, it follows that the total space of densities and elastic moduli of these HEAs are bounded by elemental properties (Fig. 8). The particular path through this elastic modulus-density space taken by a HEA reflects the compositional weights of the elemental properties, but, in general, any values within the bounds of the elemental properties should be possible if one could stabilize the compound and, as explored below, the assumptions underpinning the rule of mixtures hold. Therefore, there is not a simple relationship between elastic moduli and density owing to the large compositional space of HEAs. We note this can be used as a potentially potent design principle because it enables optimizing two commonly antagonistic properties like increasing shear modulus while lowering density.

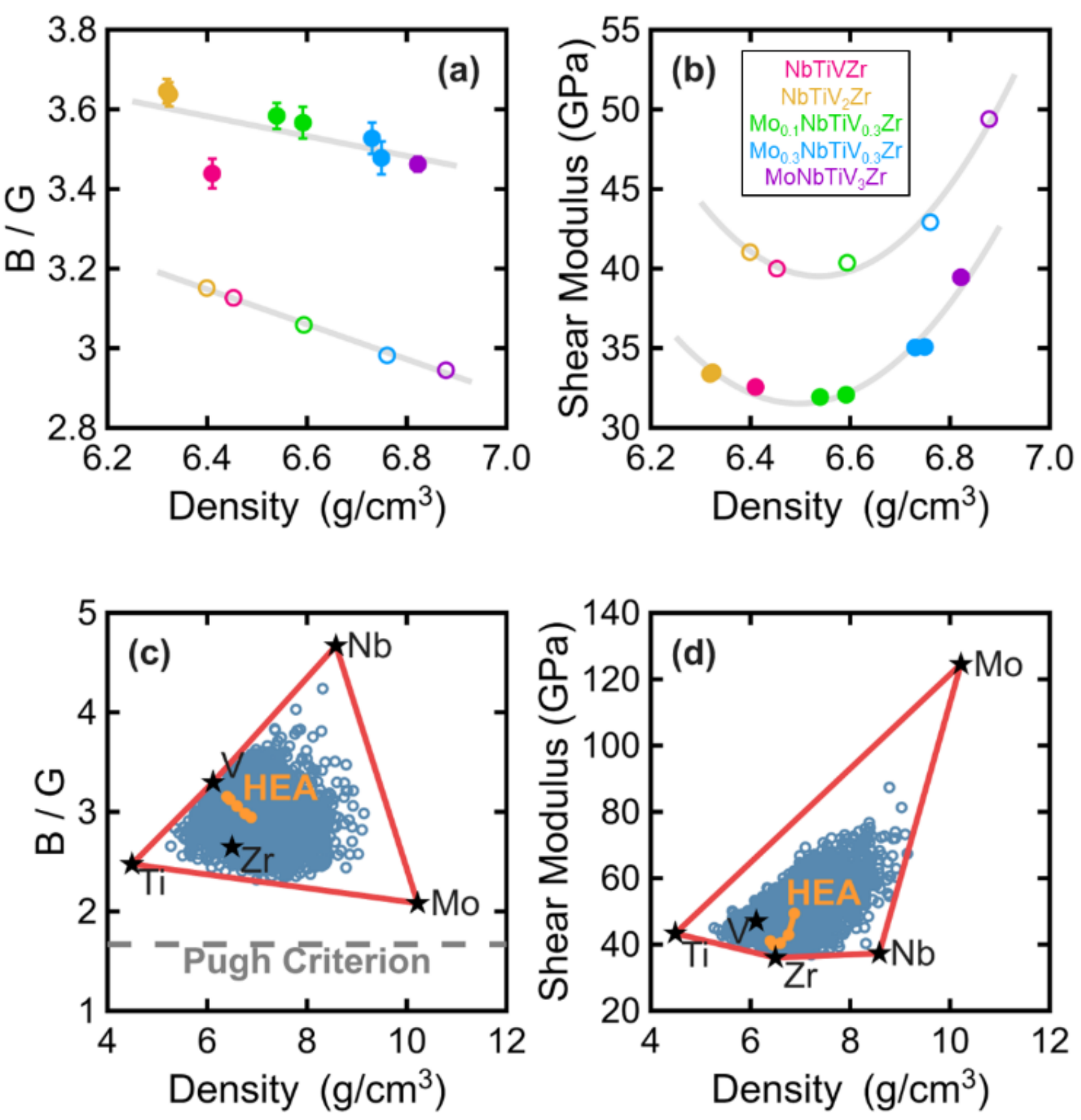

**Figure 8**. Changes in (a) the ratio of the bulk and shear moduli (an indication of intrinsic ductility and hardness) and (b) the shear moduli (an indication of strength) with density. Closed symbols are measured values and open symbols are rule of mixtures predictions using Eq. 1-3 in the text. Colors are different compositions, see inset of (b). Grey lines are visual guides. The relationship between these two properties and density is a consequence of the rule of mixtures. Black stars represent the (c) ratio of bulk and shear moduli and (d) shear moduli of elements within the Mo-Nb-Ti-V-Zr high-entropy alloy family. The area bounded by the red lines contains all possible property combinations within the rule of mixtures framework. Blue circles are simulated properties for random compositions spanning Mo=[0,1], Nb=[0,1], Ti=[0,1], V=[0,3], Zr=[0,1]. Orange circles and lines are the rule of mixtures predictions for the compositions studied in this work.

There has been much effort to predict HEA stability on the basis of their constitutive elements (*88-90*). In this spirit, the analysis presented in this work shows that the rule of mixtures framework can provide useful bounds for elastic properties in HEAs and explain experimental observations such as the counter-intuitive relationships between elastic moduli and density (Fig. 8). However, it is worth emphasizing that while the rule of mixtures can serve as guide for designing new HEAs and anticipating their properties, the quantitative differences between the rule of mixtures predictions and experimental elastic moduli (*e.g.*, Fig. 6-8) shows that experiments are ultimately needed to quantify properties. This is especially true if the core assumption underpinning the rule of mixtures (*i.e.*, a compound can be approximated as a volumetric average of each elemental constituent (see Eq. 1-3)) does not apply. The rule of mixtures is an idealized limit which estimates intrinsic contributions to elasticity arising from interatomic forces. This approximation is expected to hold when there is no significant rearrangement of charge density in going from the constituent elements of a compound to the compound itself. The rule of mixtures will provide unreliable estimates of physical properties when this is not the case and cannot capture extrinsic contributions to the effective elastic properties of a sample such as those coming from point defects, secondary phases, and cracks. For example, intermetallics, including those within the compositional space studied in this work like Nb-Ti (*91*), are known to exhibit strong deviations from rule of mixtures elastic constant predictions because bonding in intermetallics is often qualitatively different than in the elements of which they are comprised (*92*). The general agreement between the experimental and rule of mixtures bulk moduli in Mo-Nb-Ti-V-Zr refractory HEAs (Fig. 7) suggests similarity between the local bonding environments in these alloys and their constituent elements, although the differences in the shear moduli indicate differences in the angular character of the bonding in the alloys (*50-53*). Additional experiments, especially on HEA single crystals, are required for a more detailed understanding.

## IV. Conclusion

Monitoring changes in mechanical resonance frequencies and widths from variations in sample manufacturing and/or processing conditions provides valuable information on the number, type, and evolution of defects in novel materials with a relatively simple apparatus that is compatible with many existing workflows. Such measurements are fast (~minutes), precise (better than ppm), sensitive (*e.g.*, to defects and cracks), and work with any arbitrarily-shaped sample. This utility naturally pairs with the most common application of resonant ultrasound measurements, the non-destructive determination of elastic constants, to anticipate the mechanical properties of new materials and benchmark theoretical models. By simultaneously enabling a (1) rapid assessment of sample quality, (2) non-destructive determination of elastic properties, and (3) direct evaluation of theoretical models, mechanical resonance experiments are an enticing addition to the conventional materials design workflow as a platform for the design and down-selection of new materials. We anticipate this will be especially useful for systems with large parameter spaces, such as high-entropy materials.

## Materials and Methods

*Refractory High-Entropy Alloy Samples*

The arc-melted W-Ta-Cr-V-Hf and Mo-Nb-Ti-V-Zr samples were purchased from Sophisticated Alloys Inc, while the hot pressed (hot-pressed at 1500 °C) samples were purchased from MSE Supply, LLC.

*Resonant Ultrasound Spectroscopy*

Mechanical resonances of W-Ta-Cr-V-Hf and Mo-Nb-Ti-V-Zr samples at room temperature were measured using custom-built transducers (*93*) and electronics similar to those described in Ref. (*33*). Measurements at $T = 2$K were performed on a custom-built RUS probe (*94*) compatible with a Quantum Design Physical Property Measurement System (PPMS). Resonances were fit with modified Lorentzian functions from which resonant frequencies and widths were obtained (*33*). For quantitative RUS analyzes, a compositional series of Mo-Nb-Ti-V-Zr refractory high-entropy alloys were prepared into cylinders with a nominal diameter and height of 5 mm using electrical discharge machining (EDM). The dimensions (measured by micrometer) and mass were used to determine the geometric densities (see SI Table SII). Density uncertainties were found by propagating the uncertainties in the mass and volume measurements. Pycnometry corroborated the geometric densities (differences between the two density measurements were < 0.5%), indicating minimal porosity or deviations from a cylindrical shape.

Elastic constants of the Mo-Nb-Ti-V-Zr samples were determined from their resonant frequencies using the Visscher approach (*36*). The root mean square difference between the experimentally measured and predicted resonant frequencies from anisotropic elastic model was $\lesssim$ 0.4% for all samples (*18*). An example of a complete inversion result is provided in the Supporting Information. The uncertainties in the experimental elastic moduli reported throughout this work correspond to a 2% change in the $\chi^2$ value from the inversion (*18*). No thermal expansion/contraction corrections were applied for the 2 K measurements.

## Acknowledgements

Research presented in this article was supported by the Laboratory Directed Research and Development program of Los Alamos National Laboratory under project numbers 20220727ER and 20240225ER. A portion of this work was performed at the National High Magnetic Field Laboratory, which is supported by the National Science Foundation Cooperative Agreement No. DMR-2128556, the State of Florida, and the Department of Energy. This work was also supported by the U.S. Department of Energy, Advanced Research Projects Agency-Energy under contract DE-AR0001541 as well as the U.S. Department of Energy, Fusion Energy Sciences program under project No. 82396A, contract No. AT2030110.

## Conflict of Interest

The authors declare no conflict of interest.